\documentclass[preprints,article,accept,moreauthors,pdftex]{Definitions/mdpi}
\usepackage{subcaption}
\usepackage[abbreviations]{glossaries-extra}

\usepackage{mathtools}
\usepackage{soul,color}
\soulregister\cite7
\soulregister\ref7
\soulregister\pageref7
\soulregister\gls7
\soulregister&7
\sethlcolor{yellow}

\firstpage{1} 
\pubvolume{xx}
\issuenum{1}
\articlenumber{5}
\pubyear{2019}
\copyrightyear{2019}
\history{Received: date; Accepted: date; Published: date}

\Title{Lattice Boltzmann solver for multi-phase flows: Application to high Weber and Reynolds numbers}

\Author{S.A. Hosseini $^{1,2}$*, H. Safari $^{1}$ and D. Th\'evenin $^{1}$}

\AuthorNames{S.A. Hosseini, H. Safari and D. Th\'evenin}

\address{%
$^{1}$ \quad Laboratory of Fluid Dynamics and Technical Flows, University of Magdeburg ``Otto von Guericke'', D-39106 Magdeburg, Germany.\\
$^{2}$ \quad Department of Mechanical and Process Engineering, ETH Z\"urich, 8092 Z\"urich, Switzerland.}
\corres{Correspondence: seyed.hosseini@ovgu.de}

\abstract{The lattice Boltzmann method, now widely used for a variety of applications, has also been extended to model multi-phase flows through different formulations. While already applied to many different configurations in the low Weber and Reynolds number regimes, applications to higher Weber/Reynolds numbers or larger density/viscosity ratios are still the topic of active research. In this study, through a combination of the decoupled phase-field formulation --conservative Allen-Cahn equation-- and a cumulants-based collision operator for a low-Mach pressure-based flow solver, we present an algorithm that can be used for higher Reynolds/Weber numbers. The algorithm is validated through a variety of test-cases, starting with the Rayleigh-Taylor instability both in 2-D and 3-D, followed by the impact of a droplet on a liquid sheet. In all simulations, the solver is shown to correctly capture the dynamics of the flow and match reference results very well. As the final test-case, the solver is used to model droplet splashing on a thin liquid sheet in 3-D with a density ratio of 1000 and kinematic viscosity ratio of 15 --matching the water/air system-- at We=8000 and Re=1000. The results show that the solver correctly captures the fingering instabilities at the crown rim and their subsequent breakup, in agreement with experimental and numerical observations reported in the literature.}


\begin{document}

\newabbreviation{lb}{LB}{lattice Boltzmann}
\newabbreviation{lbm}{LBM}{lattice Boltzmann method}
\newabbreviation{ac}{AC}{Allen-Cahn}
\newabbreviation{ch}{CH}{Cahn-Hilliard}
\newabbreviation{pde}{PDE}{partial differential equation}
\newabbreviation{edf}{EDF}{equilibrium distribution function}
\newabbreviation{dvbe}{DVBE}{discrete velocity Boltzmann equation}
\newabbreviation{ns}{NS}{Navier-Stokes}
\newabbreviation{acm}{ACM}{artificial compressibility method}
\newabbreviation{eos}{EoS}{equation of state}
\newabbreviation{mrt}{MRT}{multiple relaxation time}
\newabbreviation{srt}{SRT}{single relaxation time}
\newabbreviation{rhs}{RHS}{right hand side}
\newabbreviation{df}{DF}{distribution function}
\newabbreviation{ce}{CE}{Chapman Enskog}

\section{Introduction}
The \gls{lbm} is a discrete solver for the so-called \gls{dvbe}, initially developed as an alternative to classical solvers for the incompressible hydrodynamic regime~\cite{kruger_lattice_2017,guo_lattice_2013}. Due to the simplicity of the algorithm, low computational cost of discrete time-evolution equations and locality of non-linear terms and boundary conditions it has rapidly grown over the past decades~\cite{succi_lattice_2002}. It is worth noting that while intended for the incompressible regime, the \gls{lbm} formally solves the compressible isothermal \gls{ns} equations at a reference temperature. While originally tied to the considered flow's temperature, in the context of the \gls{lb} solver, the reference temperature is a numerical parameter allowing to control convergence and consistency of the results~\cite{kruger_lattice_2017}. The weak compressibility in the formulation along with the parabolic nature of the \gls{pde} governing the evolution of pressure --as opposed to Chorin's original \gls{acm}-- have made the scheme efficient and applicable to unsteady flows~\cite{chorin_numerical_1997}. Although originally used for single-phase flows it has since been extended to multi-phase, multi-species and compressible flows.
\par While generally based on diffuse-interface formulations, \gls{lb} solvers for multi-phase flows can be categorized as pertaining to one of three major categories: (a) pseudo-potential~\cite{shan_lattice_1993,shan_simulation_1994}, (b) free energy~\cite{swift_lattice_1996,swift_lattice_1995} and (c) phase-field. Other types of formulations can also be found in the literature, however they are not as widely spread and/or developed as these three.\\
In the context of the free energy formulation, the expression for the non-local non-ideal pressure tensor is found through the free energy functional. The appropriate pressure tensor is then introduced into the \gls{lb} solver via a moment-matching approach assigning coefficients to different terms in the \gls{edf}~\cite{swift_lattice_1995}. The interesting point that makes this formulation consistent and differentiates it from the generic double-well potential-based Cahn-Hilliard formulation, is that in the minimization process of the free energy, the \gls{eos} is explicitly considered. It is interesting to note that, as is the case for the pseudo-potential formulation, the explicit intervention of the \gls{eos} within the free functional makes the thickness of the interface tied to physical parameters, {i.e.} surface tension, density ratio, \gls{eos} etc. As a consequence, the choice of the \gls{eos} and/or tuning of the coefficients in the \gls{eos} is a method of choice to widen the area of accessible density ratios. This approach was later extended by introducing non-ideal components of the pressure tensor via an external body force. Introducing these effects with a body force made the scheme more stable by reducing Galilean invariance issues tied to the third-order moments of the \gls{edf}~\cite{wagner_investigation_2006}.\\
The pseudo-potential formulation follows more of a bottom-up approach in introducing non-ideal dynamics into the solver. It follows the general philosophy of the Boltzmann-Vlasov equation, introducing a non-local \emph{potential} to account for non-ideal effects. While the original formulation relied on what was termed an effective \emph{density}, actual \gls{eos} were introduced into the pseudo-potential in~\cite{kupershtokh_equations_2009,yuan_equations_2006}. Apart from thermodynamic consistency, the possibility of using different \gls{eos} allowed for higher density ratios to be modeled. As the free energy formulation, this model is limited to lower Weber number regimes because it naturally comes with large surface tension values. While more advanced models allow for independent tuning of the surface tension~\cite{sbragaglia_generalized_2007}, the spectrum of values covered by the model is rather limited and barely allows for variations of one order of magnitude~\cite{li_achieving_2013}.\\
The last category is based on the free energy functional minimization approach, just like the free energy approach. However, contrary to the latter, the surface and bulk energies used in the minimization process are those of a generic double-well potential~\cite{fakhari2010phase}, allowing to decouple --among other parameters-- the interface thickness from the fluid physical properties. Another consequence of this choice of functional is a partial loss of thermodynamic consistency, making the extension of the formulation to more complex physics such as thermal flows, compressible flows, or acoustics less straightforward, although a number of attempts have been documented in the literature~\cite{safari2013extended,safari2014consistent,yazdi2018numerical}. Nevertheless, it has been observed to be very effective and robust for multi-phase flows in the incompressible regime and readily able to deal with larger Weber numbers. {For a more comprehensive overview of the developments of such models, interested readers are referred to~\cite{wang2019brief}.} It is also worth noting that approaches relying on explicit tracking of the interface with a consistent energy functional making use of the non-ideal \gls{eos} have also been proposed as ways to improve the stability of the original free energy formulation~\cite{he_lattice_1999,inamuro_lattice_2004}.
\par Over the past decades a lot of efforts have been put in developing phase field-based \gls{lb} solvers for various applications~\cite{safari2014consistent,amirshaghaghi2016application,amirshaghaghi2018large}. Given that in such formulations the local density is a dependent variable --on the local value of the order parameter, they have to be coupled to a modified form of the \gls{lb} solver for the flow usually referred to as the incompressible formulation. The so-called low-Mach formulation is mostly based on the modified distribution function introduced in~\cite{he_lattice_1999} where the pressure is the zeroth-order moment of the distribution function. This flow solver has been combined with different forms of interface tracking formulations, {e.g.} \gls{ac}, conservative \gls{ac} or \gls{ch} to model multi-phase flows. The aim of the present study is to introduce a multi-phase solver relying on the pressure-based formulation of~\cite{he_lattice_1999} and a \gls{mrt} realization --for the flow solver-- coupled with a \gls{lb} solver for the conservative \gls{ac}. The use of the \gls{mrt} collision operator in cumulants space along with the decoupled interface tracking allow for simulations in the high Reynolds and Weber regimes. After a brief introduction of the model, it will be used to simulate a variety of test-cases proving its ability to reproduce correct physics and its robustness. It is worth noting that all models were implemented in our in-house multi-physics solver ALBORZ~\cite{hosseini_development_2020}.
\section{Theoretical background}
\subsection{Target macrosopic system}
{As briefly stated in the introduction, the aim of the present work is to solve the multi-phase flow equations within the context of a diffuse interface formulation in the limit of the incompressible regime, where interface dynamics are followed and accounted for via an additional indicator field, $\phi$. As such, at the macroscopic level the low Mach \gls{ns} equations are targeted:}
\begin{equation}\label{Eq:NSEq}
{
    \partial_t \rho u_i + \partial_j \rho u_i u_j + \partial_j \sigma_{ij} + \mu_{\phi} \partial_i \phi + F_{b,i} = 0,}
\end{equation}
{where $u_i$ is the fluid velocity, $\rho$ the fluid density and $F_{b,i}$ designates external body forces. The stress tensor $\sigma_{ij}$ is defined as:}
\begin{equation}
{
    \sigma_{ij} = p_h \delta_{ij} - \eta \left( \partial_i u_j + \partial_j u_i\right) + \frac{2}{3}\eta \partial_k u_k \delta_{ij},}
\end{equation}
{where $\eta$ is the fluid dynamic viscosity, tied to the kinematic viscosity $\nu$ as $\eta=\rho\nu$, and $p_h$ the hydrodynamic pressure. The chemical potential $\mu_{\phi}$ is defined as:}
\begin{equation}
{
    \mu_{\phi} = 2\beta\phi\left(\phi-1\right)\left(2\phi-1\right) \kappa \Delta \phi,}
\end{equation}
{where $\Delta=\nabla^2$ is the Laplacian operator and $\beta$ and $\kappa$ are parameters specific to the \gls{ac} formulation. It must be noted that the second term on the \gls{rhs} of Equation~\ref{Eq:NSEq} accounts for surface tension effects. For the sake of clarity the free parameters will be detailed in the next paragraph.}\\
{The interface is tracked using the conservative \gls{ac} equation, where the order parameter $\phi$ evolves as~\cite{sun2007sharp,chiu2011conservative}:}
\begin{equation}\label{eq:AC_macro}
{
    \partial_t \phi + \partial_i u_i \phi - \partial_i M \left[ \partial_i \phi - n_i \frac{4\phi(1 - \phi)}{W}\right] = 0,}
\end{equation}
{where the parameter $\phi$ takes on values between 0 and 1, $M$ is mobility, $W$ is the interface thickness and $n_i$ is the unit normal to the interface obtained as:}
\begin{equation}
{
    n_i = \frac{\partial_i \phi}{\lvert\lvert \nabla\phi\lvert\lvert}.}
\end{equation}
{The interfaces can be found through iso-surfaces of the order parameter, {i.e.} $\phi=1/2$. To recover the correct surface tension, the free parameters appearing in the chemical potential, {i.e.} $\kappa$ and $\beta$ are tied to the surface tension $\sigma$ and interface thickness $W$ in the \gls{ac} equation via $\beta=12\sigma/W$ and $\kappa=3\sigma W/2$.}
\subsection{\gls{lb} formulation for the conservative phase-field equation}
The conservative \gls{ac} equation can readily be recovered by appropriately defining the discrete equilibrium state and relaxation coefficient in the advection-diffusion \gls{lb} model:
\begin{equation}\label{eq:LBM_AC}
    \partial_t g_\alpha + c_{\alpha,i}\partial_i g_\alpha + \mathcal{S}_\alpha = \Omega^{\phi}_\alpha,
\end{equation}
where $g_\alpha$ and $c_\alpha$ are the populations and velocities in the discrete velocity kinetic model and the collision operator is defined as:
\begin{equation}
    \Omega^{\phi}_\alpha = \frac{1}{\tau_{\phi}}\left(g_\alpha^{(eq)}-g_\alpha\right).
\end{equation}
The \gls{edf} is defined as:
\begin{equation}
{
    g_\alpha^{(eq)} = w_\alpha \phi \sum_{n=0}^2 \frac{1}{n!c_s^{2n}}\mathcal{H}_n:a^{(eq)}_n,
    }
\end{equation}
where $\mathcal{H}_n$ and $a_n^{(eq)}$ are the Hermite polynomial and coefficient of order $n$, $c_s$ the lattice sound speed and $w_\alpha$ weights tied to each discrete velocity (resulting from the Gauss-Hermite quadrature). The expressions for these polynomials and corresponding coefficients are listed in Appendix~\ref{App:Hermite}. The source term in Equation~\ref{eq:LBM_AC} is defined as~\cite{fakhari_weighted_2017}:
\begin{equation}
    \mathcal{S}_\alpha = w_\alpha \mathcal{H}_i n_i \frac{4\phi(1-\phi)}{W}.
\end{equation}
Given that the source term affects the first-order moment --a non-conserved moment of the distribution function-- the distribution function is tied to the phase parameter as:
\begin{equation}
    \phi = \sum_\alpha g_\alpha.
\end{equation}
The relaxation coefficient is fixed as:
\begin{equation}
    \tau_\phi = \frac{M}{c_s^2}.
\end{equation}
{After integration in space/time the now-famous collision-streaming form can be recovered:}
\begin{equation}
{
    \bar{g}_\alpha\left(x+c_\alpha \delta_t, t+\delta_t\right) = \left(1-\frac{\delta_t}{\bar{\tau}_{\phi}}\right)\bar{g}_\alpha\left(x, t\right) + \frac{\delta_t}{\bar{\tau}_{\phi}}g^{(eq)}_\alpha\left(x, t\right) + \delta_t\bar{\mathcal{S}}_\alpha\left(x, t\right),
    }
\end{equation}
{where the source term takes on a new form, {i.e.}:}
\begin{equation}\label{eq:phase_source_post_discretization}
    \bar{\mathcal{S}}_\alpha = \left(1-\frac{1}{2\tau_\phi}\right)w_\alpha \mathcal{H}_i n_i \frac{4\phi(1-\phi)}{W},
\end{equation}
{and:}
\begin{equation}
{
    \bar{\tau}_\phi = \tau_\phi + \frac{\delta_t}{2}.
    }
\end{equation}
It is also worth noting that the derivatives of the order parameter appearing in the various discrete time-evolution equations are computed using isotropic finite differences, {i.e.}:
\begin{equation}
    \partial_i \phi = \frac{1}{c_s^2}\sum_\alpha w_\alpha c_{\alpha,i} \phi(x+c_\alpha),
\end{equation}
and:
\begin{equation}
    \partial^2_i \phi = \frac{2}{c_s^2}\sum_\alpha w_\alpha \left[\phi(x+c_\alpha) - \phi(x)\right].
\end{equation}
{While the present work makes use of a second-order \gls{edf}, one must note that the same macroscopic \gls{pde}, {i.e.} Equation~\ref{eq:AC_macro}, }can also be recovered by using a first-order \gls{edf} and an additional correction term of the following form~\cite{wang2016comparative}:
\begin{equation}
    C_\alpha = \frac{w_\alpha}{c_s^2} \mathcal{H}_i\partial_t \phi u_i,
\end{equation}
where, as for Equation~\ref{eq:phase_source_post_discretization}, post-discretization it changes into:
\begin{equation}
    \bar{C}_\alpha = \left(1-\frac{1}{2\tau_\phi}\right)\frac{w_\alpha}{c_s^2} \mathcal{H}_i\partial_t \phi u_i.
\end{equation}
Such correction terms were first introduced in the context of advection-diffusion \gls{lb} solvers~\cite{chopard2009lattice} and further extended to non-linear equations in the same context~\cite{hosseini2019lattice}. {Detailed derivation and multi-scale analyses are readily available in the literature, {e.g.}~\cite{zu2020phase}.}
\subsection{\gls{lb} model for flow field}
The flow solver kinetic model follows the low-Mach formulation used, among other sources, in~\cite{lee2003pressure,hosseini2019hybrid,hosseini2020low} and based on the original model introduced in~\cite{he_lattice_1999}:
\begin{equation}\label{eq:LMNA_LBM}
    \partial_t f^{'}_\alpha + c_{\alpha,i}\partial_i f^{'}_\alpha = \Omega_\alpha + \Xi_\alpha,
\end{equation}
where the collision operator is:
\begin{equation}\label{eq:LMNA_collision}
    \Omega_\alpha = \frac{1}{\tau}\left({f_\alpha^{(eq)}}^{'} - f^{'}_\alpha\right),
\end{equation}
and $\Xi_\alpha$ is defined as:
\begin{equation}\label{eq:LMNA_source}
    \Xi_\alpha =  c_s^2\left(\frac{f_\alpha^{(eq)}}{\rho}-w_\alpha \right)\left(c_{\alpha,i}-u_i\right)\partial_i \rho + w_\alpha c_s^2 \rho\partial_i u_i + \left(F_{b,i}+F_{s,i}\right)\left(c_{\alpha,i} - u_i\right)\frac{f_\alpha^{(eq)}}{\rho},
\end{equation}
and the relaxation coefficient $\tau$ is tied to the fluid kinematic viscosity $\nu$ as:
\begin{equation}\label{eq:LMNA_relaxation_coefficient}
    \tau = \frac{\nu}{c_s^2}.
\end{equation}
The forces $F_{b,i}$ and $F_{s,i}$ represent respectively external body forces and surface tension, {i.e.}:
\begin{equation}
{
    F_{s,i} = \mu_{\phi} \partial_i \phi.
    }
\end{equation}
The modified distribution function, $f^{'}_\alpha$, is defined as:
\begin{equation}\label{eq:LMNA_distribution}
    f^{'}_\alpha = w_\alpha p_h + c_s^2 \left(f_\alpha - w_\alpha \rho \right),
\end{equation}
where $f_\alpha$ is the classical iso-thermal distribution function. {The modified equilibrium follows the same logic and is defined as:}
\begin{equation}
{
    {f^{(eq)}_\alpha}^{'} = w_\alpha p_h + w_\alpha \rho c_s^2\sum_{n=1}^{2} \frac{1}{n!c_s^{2n}} \mathcal{H}_n:a^{(eq)}_n.}
\end{equation}
The density is tied to the order parameter as:
\begin{equation}
    \rho = \rho_l + \left(\rho_h - \rho_l\right) \phi,
\end{equation}
where $\rho_h$ and $\rho_l$ are respectively the densities of the heavy and light fluid. {For a detailed analysis of the macroscopic equations recovered by this model and the derivation of the discrete equations, interested readers are referred to~\cite{hosseini2019hybrid,hosseini_development_2020}.}
In the context of the present study the low-Mach model is wrapped in a moments-based formulation where the post-collision populations $f^{'*}_\alpha$, {to be streamed as:}
\begin{equation}
{
    {f^{'}}_\alpha\left(x+c_\alpha \delta_t, t+\delta_t\right) = {f^{'*}}_\alpha\left(x, t\right),}
\end{equation}
are computed as:
\begin{equation}
    {f^{'*}}_\alpha = \rho c_s^2 f_\alpha^{p*} + \frac{\delta_t}{2}\Xi_\alpha.
\end{equation}
The post-collision pre-conditioned population $f_\alpha^{p*}$ is:
\begin{equation}
     {f^{p*}_\alpha} = \mathcal{C}^{-1}\left(\mathcal{I} - \mathcal{W}\right)\mathcal{K}^{p} + \mathcal{C}^{-1}\mathcal{W}\mathcal{K}^{p},
\end{equation}
where $\mathcal{C}$ is the moments transform matrix --from pre-conditioned populations to the target momentum space, $\mathcal{I}$ the identity matrix and $\mathcal{W}$ the diagonal relaxation frequency matrix. Following~\cite{geier_under-resolved_2020} prior to transformation to momentum space the populations are pre-conditioned as:
\begin{equation}
    f^{p}_\alpha = \frac{1}{\rho c_s^2} f^{'}_\alpha + \frac{\delta_t}{2\rho c_s^2}\Xi_\alpha.
\end{equation}
This pre-conditioning accomplishes two tasks, namely normalizing the populations with the density and thus eliminating the density-dependence of the moments and introducing the first half of the source term. As such the moments $\mathcal{K}^{p}$ are computed as:
\begin{equation} \label{eq:C_alpha_beta}
    \mathcal{K}^{p}_{\beta} = \mathcal{C}_{\alpha\beta} f^{p}_\alpha.
\end{equation}
{The transformation from \gls{df}s to cumulants is carried out using the steps suggested in~\cite{geier_cumulant_2015}, which allows for a more efficient algorithm. The \gls{df}s are first transformed into central moments:}
\begin{equation} \label{eq:df_to_central}
{
    \widetilde{\Pi}_\beta^{p} = \sum_\alpha {\left(c_{\alpha,x}-u_x\right)}^{n_x} {\left(c_{\alpha,y}-u_y\right)}^{n_y} {\left(c_{\alpha,z} - u_z\right)}^{n_z} f_\alpha^{p}.
    }
\end{equation}
{where here $\beta=x^{n_x}y^{n_y}z^{n_z}$. The central moments are then transformed into the corresponding cumulants using the following relations:}
\begin{subequations}
\begin{align}
    \mathcal{K}^{p}_{x} = &\widetilde{\Pi}_{x}^{p}, \\
    \mathcal{K}^{p}_{xy} = &\widetilde{\Pi}_{xy}^{p}, \\
    \mathcal{K}^{p}_{x^2} = &\widetilde{\Pi}_{x^2}^{p}, \\
    \mathcal{K}^{p}_{xy^2} = &\widetilde{\Pi}_{xy^2}^{p}, \\
    \mathcal{K}^{p}_{xyz} = &\widetilde{\Pi}_{xyz}^{p}, \\
    \mathcal{K}^{p}_{x^2yz} = &\widetilde{\Pi}_{x^2yz}^{p} - \left[\widetilde{\Pi}_{x^2}^{p}\widetilde{\Pi}_{yz}^{p} + 2\widetilde{\Pi}_{xy}^{p}\widetilde{\Pi}_{xz}^{p}\right], \\
    \mathcal{K}^{p}_{x^2y^2} = &\widetilde{\Pi}_{x^2y^2}^{p} - \left[\widetilde{\Pi}_{x^2}^{p}\widetilde{\Pi}_{y^2}^{p} + 2{(\widetilde{\Pi}_{xy}^{p})}^2\right], \\
    \mathcal{K}^{p}_{xy^2z^2} = &\widetilde{\Pi}_{xy^2z^2}^{p} - \left[\widetilde{\Pi}_{z^2}^{p}\widetilde{\Pi}_{xy^2}^{p} + \widetilde{\Pi}_{y^2}^{p}\widetilde{\Pi}_{xz^2}^{p} + 4\widetilde{\Pi}_{yz}^{p}\widetilde{\Pi}_{xyz}^{p} + 2(\widetilde{\Pi}_{xz}^{p}\widetilde{\Pi}_{y^2z}^{p} + \widetilde{\Pi}_{xy}^{p}\widetilde{\Pi}_{yz^2}^{p})\right], \\
    \mathcal{K}^{p}_{x^2y^2z^2} = &\widetilde{\Pi}_{x^2y^2z^2}^{p} - \left[4{(\widetilde{\Pi}_{xyz}^{p})}^2 + \widetilde{\Pi}_{x^2}^{p}\widetilde{\Pi}_{y^2z^2}^{p} + \widetilde{\Pi}_{y^2}^{p}\widetilde{\Pi}_{x^2z^2}^{p} + \widetilde{\Pi}_{z^2}^{p}\widetilde{\Pi}_{x^2y^2}^{p} + 4(\widetilde{\Pi}_{xy}^{p}\widetilde{\Pi}_{x^2yz}^{p} + \right. \nonumber \\
    & \left. \widetilde{\Pi}_{xz}^{p}\widetilde{\Pi}_{xy^2z}^{p} + \widetilde{\Pi}_{xy}^{p}\widetilde{\Pi}_{xyz^2}^{p} +2(\widetilde{\Pi}_{xy^2}^{p}\widetilde{\Pi}_{xz^2}^{p} + \widetilde{\Pi}_{x^2y}^{p}\widetilde{\Pi}_{yz^2}^{p} + \widetilde{\Pi}_{x^2z}^{p}\widetilde{\Pi}_{y^2z}^{p})) + \right. \nonumber \\
    & \left. (16\widetilde{\Pi}_{xy}^{p}\widetilde{\Pi}_{xz}^{p}\widetilde{\Pi}_{yz}^{p} + 4({(\widetilde{\Pi}_{xz}^{p})}^2\widetilde{\Pi}_{y^2}^{p} + {(\widetilde{\Pi}_{yz}^{p})}^2\widetilde{\Pi}_{x^2}^{p} + {(\widetilde{\Pi}_{xy}^{p})}^2\widetilde{\Pi}_{z^2}^{p}) + 2\widetilde{\Pi}_{x^2}^{p}\widetilde{\Pi}_{y^2}^{p}\widetilde{\Pi}_{z^2}^{p})\right].
\end{align}
\end{subequations}
{The remainder of the moments can be easily obtained via permutation of the indices. The collision process is performed in cumulant space according to ~\cite{geier_cumulant_2015}. The fluid viscosity is controlled via the collision factor related to second-order cumulants ({e.g.} $\mathcal{K}^{p}_{xy}$, $\mathcal{K}^{p}_{x^2} - \mathcal{K}^{p}_{y^2}$, $\mathcal{K}^{p}_{x^2} - \mathcal{K}^{p}_{z^2}$ etc). The rest of the collision factors are set to unity for simplicity. Once the collision step has been applied, cumulants are transformed back into central moments as:}
\begin{subequations}
\begin{align}
    \widetilde{\Pi}_{x}^{p*} =& \mathcal{K}^{p*}_{x}, \\
    \widetilde{\Pi}_{xy}^{p*} =& \mathcal{K}^{p*}_{xy}, \\
    \widetilde{\Pi}_{x^2}^{p*} =& \mathcal{K}^{p*}_{x^2}, \\
    \widetilde{\Pi}_{xy^2}^{p*} =& \mathcal{K}^{p*}_{xy^2}, \\
    \widetilde{\Pi}_{xyz}^{p*} =& \mathcal{K}^{p*}_{xyz}, \\
    \widetilde{\Pi}_{x^2yz}^{p*} = & \mathcal{K}^{p*}_{x^2yz} + \left[\widetilde{\Pi}^{p*}_{x^2}\widetilde{\Pi}^{p*}_{yz} + 2\widetilde{\Pi}^{p*}_{xy}\widetilde{\Pi}^{p*}_{xz}\right], \\
    \widetilde{\Pi}_{x^2y^2}^{p*} = & \mathcal{K}^{p*}_{x^2y^2} + \left[\widetilde{\Pi}_{x^2}^{p*}\widetilde{\Pi}_{y^2}^{p*} + 2{(\widetilde{\Pi}_{xy}^{p*})}^2 \right], \\
     \widetilde{\Pi}_{xy^2z^2}^{p*} = & \mathcal{K}^{p*}_{xy^2z^2} + \left[\widetilde{\Pi}^{p*}_{z^2}\widetilde{\Pi}^{p*}_{xy^2} + \widetilde{\Pi}^{p*}_{y^2}\widetilde{\Pi}^{p*}_{xz^2} + 4\widetilde{\Pi}^{p*}_{yz}\widetilde{\Pi}^{p*}_{xyz} + 2(\widetilde{\Pi}^{p*}_{xz}\widetilde{\Pi}^{p*}_{y^2z} + \widetilde{\Pi}^{p*}_{xy}\widetilde{\Pi}^{p*}_{yz^2})\right], \\
    \widetilde{\Pi}^{p*}_{x^2y^2z^2} = & \mathcal{K}^{p*}_{x^2y^2z^2} + \left[4{(\widetilde{\Pi}^{p*}_{xyz})}^2 + \widetilde{\Pi}^{p*}_{x^2}\widetilde{\Pi}^{p*}_{y^2z^2} + \widetilde{\Pi}^{p*}_{y^2}\widetilde{\Pi}^{p*}_{x^2z^2} + \widetilde{\Pi}^{p*}_{z^2}\widetilde{\Pi}^{p*}_{x^2y^2} + 4(\widetilde{\Pi}^{p*}_{xy}\widetilde{\Pi}^{p*}_{x^2yz} + \right. \nonumber \\
     & \left. \widetilde{\Pi}^{p*}_{xz}\widetilde{\Pi}^{p*}_{xy^2z} + \widetilde{\Pi}^{p*}_{xy}\widetilde{\Pi}^{p*}_{xyz^2} +2(\widetilde{\Pi}^{p*}_{xy^2}\widetilde{\Pi}^{p*}_{xz^2} + \widetilde{\Pi}^{p*}_{x^2y}\widetilde{\Pi}^{p*}_{yz^2} + \widetilde{\Pi}^{p*}_{x^2z}\widetilde{\Pi}^{p*}_{y^2z})) - \right. \nonumber \\
     & \left. (16\widetilde{\Pi}^{p*}_{xy}\widetilde{\Pi}^{p*}_{xz}\widetilde{\Pi}^{p*}_{yz} + 4({(\widetilde{\Pi}^{p*}_{xz})}^2\widetilde{\Pi}^{p*}_{y^2} + {(\widetilde{\Pi}^{p*}_{yz})}^2\widetilde{\Pi}^{p*}_{x^2} + {(\widetilde{\Pi}^{p*}_{xy})}^2\widetilde{\Pi}^{p*}_{z^2}) + 2\widetilde{\Pi}^{p*}_{x^2}\widetilde{\Pi}^{p*}_{y^2}\widetilde{\Pi}^{p*}_{z^2})\right].
\end{align}
\end{subequations}
{After this step, the post-collision central moments can be readily transformed back to populations. All transforms presented here and upcoming simulations are based on the D3Q27 stencil. It must also be noted that the following set of 27 moments are used as moments basis:}
\begin{multline}
    \beta \in \{0, x, y, z, xy, xz, yz, x^2-y^2, x^2-z^2, x^2+y^2+z^2, \\xy^2+xz^2, xyz, xy^2-xz^2, x^2+yz^2, x^2z+y^2z, x^2y-yz^2, x^2z-y^2z, x^2y^2-2x^2z^2+y^2z^2,\\
    x^2y^2+x^2z^2-2y^2z^2, x^2y^2+x^2z^2+y^2z^2, x^2yz, xy^2z, xyz^2, x^2y^2z, x^2yz^2, xy^2z^2, x^2y^2z^2\},
\end{multline}
{where $\beta=x^2-y^2$ stands for a central moment of the form $\widetilde{\Pi}^{p}_{x^2}-\widetilde{\Pi}^{p}_{y^2}$. Previous systematic studies of the flow solver have shown second-order convergence under diffusive scaling~\cite{hosseini2019hybrid}.}
\section{Numerical applications}
The proposed numerical method will be validated through different test-cases in the present section. All results and simulation parameters are reported in \gls{lb} units, {i.e.} non-dimensionalized with time-step, grid-size and heavy fluid density.
\subsection{Static droplet: Surface tension measurement}
\begin{figure}[!ht]
	\centering
	\hspace{-0.2\textwidth}
	\begin{subfigure}{0.3\textwidth}
		\includegraphics{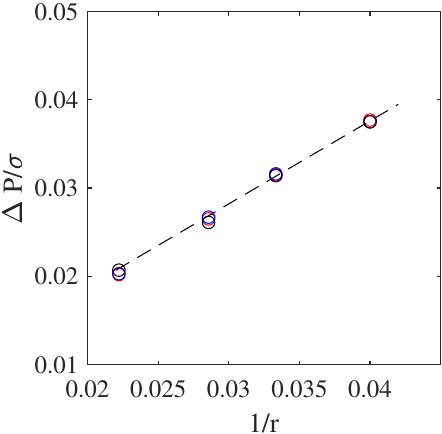}
	\end{subfigure}
	\caption{Changes of pressure difference around droplet for different surface tensions and droplet radii. {Red, blue and black symbols illustrate results from the present study with respectively $\sigma=10^{-1}, 10^{-3}$ and $10^{-6}$.}}
	\label{fig:laplace_law}
\end{figure}

As a first test, to validate the hydrodynamics of the model, we consider the case of a static droplet in a rectangular domain with periodic boundaries all around. All cases consist of a domain of size $256\times256$ filled with a light fluid. A \emph{droplet} of the heavier fluid is placed at the center of the domain. Simulations are pursued till the system converges. The pressure difference between the droplet and surrounding lighter fluid is then extracted. Using Laplace's law, {i.e.}:
\begin{equation}
    \Delta P = \frac{\sigma}{r},
\end{equation}
where $\Delta P$ is the pressure difference and $r$ the droplet radius, one can readily obtain the effective surface tension. Three different surface tensions, {i.e.} $\sigma=1\times10^{-1}$, $1\times10^{-3}$ and $1\times10^{-6}$, along with four different droplet radii, {i.e.} $r=25$, $30$, $35$ and $45$, were considered here. The obtained results are shown in Figure~\ref{fig:laplace_law}. The results presented here consider a density ratio of 20 {and non-dimensional viscosity of 0.1.}
It is readily observed that the model satisfies Laplace's law and recovers correct surface tensions. Furthermore, it is seen that it can span a wide range of surface tensions, as opposed to other classes of multi-phase solvers such as the free energy or pseudo-potential formulations~\cite{qin_entropic_2018,mazloomi_m_entropic_2015} {and maintain relatively low spurious currents. For example, at a density ratio of 1000 and $\sigma=10^{-3}$, the spurious currents were found to be only of the order of $10^{-6}$, in strong contrast with the previously cited approaches.}
\subsection{Rayleigh-Taylor instability}
\begin{figure}[!htb]
	\centering
	\hspace{-0.5\textwidth}
	\begin{subfigure}{0.3\textwidth}
		 \includegraphics{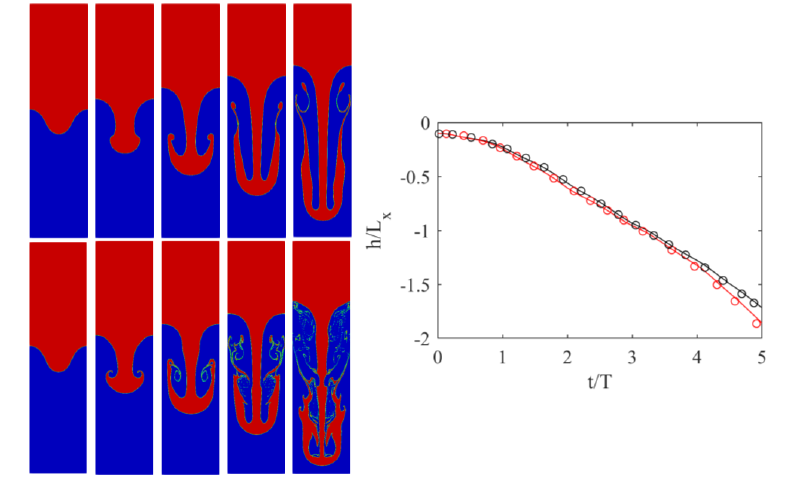}
	\end{subfigure}
\caption{{(Right) Evolution of interface for the Rayleigh-Taylor instability for (top row) Re=256 and (bottom row) Re=2048 at different times: (from left to right) $t/T=$1,2,3,4 and 5. (Left) Position of the penetrating spike over time: (black) Re=256 and (red) Re=2048. (Plain lines) present results and (symbols) data from~\cite{he_lattice_1999}.}}
\label{Fig:RTI_contours}
\end{figure}

\begin{figure}[!htb]
	\centering
	\hspace{-0.5\textwidth}
	\begin{subfigure}{0.3\textwidth}
		 \includegraphics{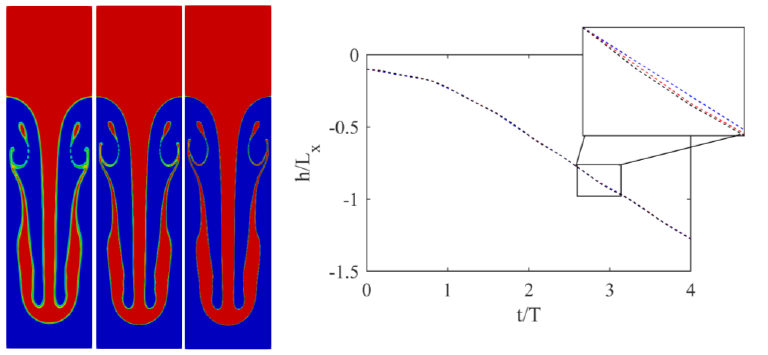}
	\end{subfigure}
\caption{{(Right) Interface for the Rayleigh-Taylor instability at $t/T=$5 and Re=256 for three different resolutions (left to right) $L_x$=150, 300 and 600. (Left) Position of the penetrating spike over time: (black) $L_x$=600, (red) $L_x$=300 and (blue) $L_x$=150.}}
\label{Fig:RTI_contours_convergence_256}
\end{figure}

\begin{figure}[!htb]
	\centering
	\hspace{-0.5\textwidth}
	\begin{subfigure}{0.3\textwidth}
		 \includegraphics{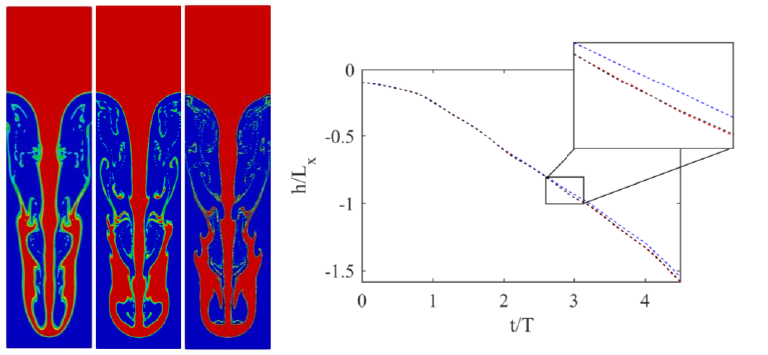}
	\end{subfigure}
\caption{{(Right) Interface for the Rayleigh-Taylor instability at $t/T=$5 and Re=2048 for three different resolutions (left to right) $L_x$=150, 300 and 600. (Left) Position of the penetrating spike over time: (black) $L_x$=600, (red) $L_x$=300 and (blue) $L_x$=150.}}
\label{Fig:RTI_contours_convergence_Re2048}
\end{figure}

The Rayleigh-Taylor instability is a well-known and widely studied gravity-driven effect occurring when a layer of a heavier fluid lies on top of another layer of a lighter fluid~\cite{yang2018entropy,yang2018statistics,rahmat2014numerical}. A perturbation at the interface between the two fluids causes the heavier one to penetrate into the lighter fluid. In general, the dynamics of this system are governed by two non-dimensional parameters, namely the Atwood and Reynolds numbers. The former is defined as:
\begin{equation}
    \hbox{At} = \frac{\rho_h - \rho_l}{\rho_h + \rho_l},
\end{equation}
while the latter is:
\begin{equation}
    \hbox{Re} = \frac{\rho_h U^{*} L}{\mu_h},
\end{equation}
where $\rho_l$ and $\rho_h$ are the densities of the heavy and light fluids, $\mu_h$ is the dynamic viscosity of the heavy fluid, $L_x$ is the size of the domain in the horizontal direction and $U^{*}$ is the characteristic velocity defined as:
\begin{equation}
    U^{*} = \sqrt{gL_x},
\end{equation}
where $g$ is gravity-driven acceleration. The characteristic time for this case is defined as:
\begin{equation}
    T = \frac{L_x}{U^{*}}.
\end{equation}
Following the set-up studied in~\cite{he_lattice_1999}, we consider a domain of size $L_x\times4L_x$ with $L_x=600$. Initially the top half of the domain is filled with the heavy liquid and the bottom half with the lighter one. The interface is perturbed via the following profile:
\begin{equation}
    h_i(x) = \frac{L}{10}\cos\left(\frac{2\pi x}{L_x}\right) + 2L_x.
\end{equation}
{While periodic boundaries were applied in the horizontal direction, at the top and bottom boundaries no-slip boundary conditions were applied using the half-way bounce-back scheme~\cite{kruger_lattice_2017}.} The At number is set to 0.5 while two different Re numbers are considered, {i.e.} Re=256 and 2048. {In both cases $g=6\times10^{-6}$ while the non-dimensional viscosities were respectively 0.1406 and 0.0176.} To validate the simulations, the position of the downward-plunging heavy liquid spike is measured over time and compared to reference data from~\cite{he_lattice_1999}. The results are illustrated in Figure~\ref{Fig:RTI_contours}.
It is observed that both simulations agree very well with the reference solution of~\cite{he_lattice_1999}. {To showcase the ability of the solver to handles under-resolved simulations and illustrate the convergence of the obtained solutions, the simulations were repeated at two additional lower resolutions with $L_x=$300 and 150 with an acoustic scaling of the time-step size. The results obtained with those lower resolutions are shown in Figures~\ref{Fig:RTI_contours_convergence_256} and \ref{Fig:RTI_contours_convergence_Re2048}.}
{By looking at the position of the plunging spike it can be clearly seen that while minor differences exist, even the lowest resolution  captures the correct position. Smaller feature however, especially at Re=2048, need higher resolutions to be correctly captured. At Re=256 for instance, even the secondary instability is converged as at $L_x$=300 no segmentation is observed. For Re=2056 on the other hand, while larger structure start to converge, thinner features clearly need more resolutions.}
\subsection{Turbulent 3-D Rayleigh-Taylor instability}
To further showcase the ability of the solver to deal with complex flows, we also consider the Rayleigh-Taylor instability in 3-D. The studied configuration follows those studied in~\cite{liang_lattice_2016}. The definitions of non-dimensional parameters are similar to those used in the previous section. The domain is discretized using $100\times100\times1200$ grid-points, with $L=100$. The interface is placed at the center of the domain along the $z$-axis, and perturbed using:
\begin{equation}
    h_i(x,y) = \frac{L}{10}\left[ \cos\left(\frac{2\pi x}{L}\right) + \cos\left(\frac{2\pi y}{L}\right) \right] + 6L,
\end{equation}
and the Reynolds and Atwood numbers are set to respectively 1000 and 0.15. As for previous configurations, {periodic boundaries were applied in the horizontal direction and no-slip boundaries at the top and bottom. The body force was set to $g=3.6\times10^{-5}$ and viscosity to 0.006.} The position of the downward-plunging spike was measured over time and compared to reference data from~\cite{liang_lattice_2016}. After the penetration of two liquids into each other, the Kelvin-Helmholtz instability causes the plunging spike to roll up and take a mushroom-like shape. As the mushroom-shaped spike further progresses into the lighter fluid, the cap disintegrates into four finger-like structures. It is interesting to note that, as will be shown later, these fingers are reminiscent of the instabilities leading to splashing in the impact of a droplet on liquid surfaces.
\begin{figure}[!htb]
	\centering
	\hspace{-0.5\textwidth}
	\begin{subfigure}{0.3\textwidth}
		 \includegraphics{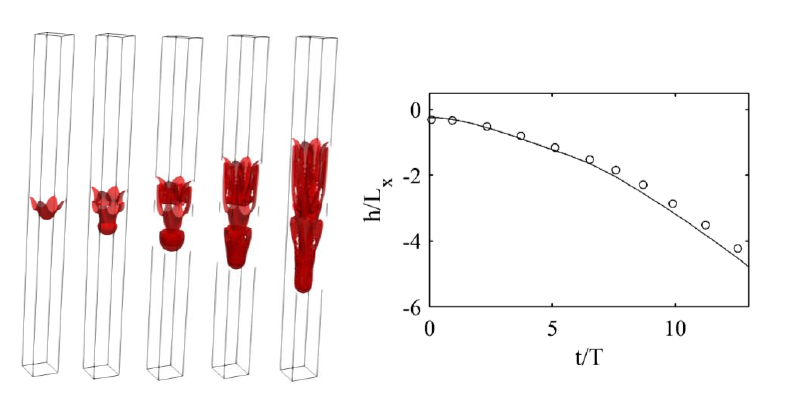}
	\end{subfigure}
\caption{(Right) Evolution of interface for the 3-D Rayleigh-Taylor instability for Re=1000 at different times: (from left to right) $t/T=$1.9, 3.9, 5.8, 7.8 and 9.7. (Left) Position of the penetrating spike over time: (Plain lines) present results and (symbols) data from~\cite{liang_lattice_2016}.}
\label{Fig:RTI3D_contours}
\end{figure}
Overall, as shown in Figure~\ref{Fig:RTI3D_contours}, the results obtained from the present simulation are in good agreement with reference data.
\subsection{Droplet splashing on thin liquid film}
\begin{figure}[!htb]
	\centering
	\hspace{-0.4\textwidth}
	\begin{subfigure}{0.3\textwidth}
		 \includegraphics{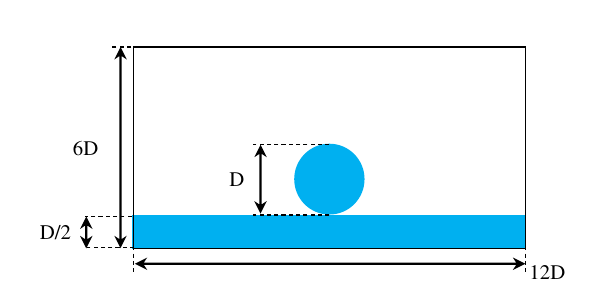}
	\end{subfigure}
\caption{Geometrical configuration of the droplet impact on liquid sheet case in 2-D.}
\label{Fig:drop_splashing_geometry}
\end{figure}

\begin{figure}[!htb]
	\centering
	\hspace{-0.55\textwidth}
	\begin{subfigure}{0.3\textwidth}
		 \includegraphics{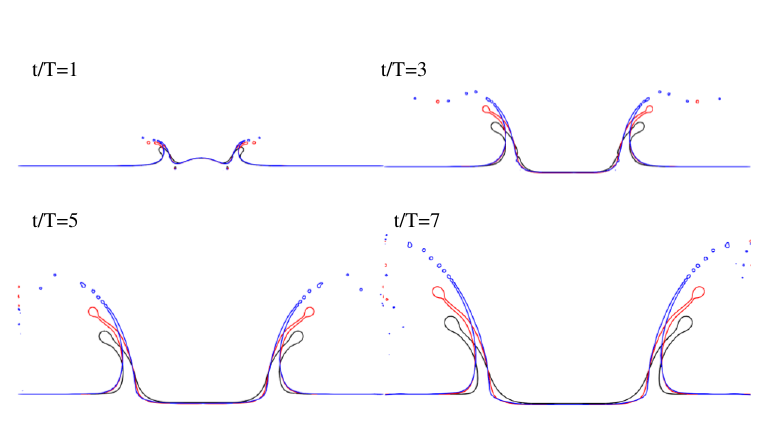}
	\end{subfigure}
\caption{Impact of circular droplet on liquid sheet at different We and Re numbers with $\rho_h/\rho_l=1000$ and $\nu_l/\nu_h=15$. (black) Re=200 and We=220, (red) Re=1000 and We=220, and (blue) Re=1000 and We=2200.}
\label{Fig:drop_splashing_2D_contours}
\end{figure}

\begin{figure}[!ht]
	\centering
	\hspace{-0.2\textwidth}
	\begin{subfigure}{0.3\textwidth}
		\includegraphics{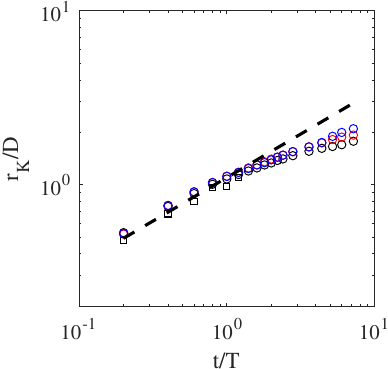}
	\end{subfigure}
	\caption{Evolution of the spreading radius $r_K$ as a function of time for the droplet impact on liquid film case. Circular symbols designate 2-D simulations: (black) Re=200 and We=220, (red) Re=1000 and We=220 and (blue) Re=1000 and We=2200. Rectangular symbols belong to the 3-D simulation with Re=1000 and We=8000. The dashed line is $\frac{r_K}{D}=1.1\sqrt{t/T}$.}
	\label{fig:drop_splashing_radius}
\end{figure}

As the final case, we consider the impact of a droplet on a thin liquid layer. This configuration is interesting as it involves complex dynamics such as splashing and is of interest in many areas of science and engineering~\cite{hagemeier2011practice,hagemeier2012experimental}. Immediately after impact the liquid surface is perturbed. In many instances, at the contact point (line) a thin liquid jet forms, and then continues to grow and propagate as a \emph{corolla}. As the crown-like structures propagates radially, a rim starts to form. At high enough Weber numbers the structure breaks into small droplets via the Rayleigh–Plateau instability~\cite{josserand_droplet_2003}. A detailed study of the initial stages of the spreading process have shown that the spreading radius scales with time regardless of the Weber and Reynolds numbers~\cite{josserand_droplet_2003}. While widely studied in the literature using different numerical formulations~\cite{hu_hybrid_2019,liang_phase-field-based_2018,fakhari_weighted_2017,sitompul_filtered_2019}, simulations have usually been limited to lower density and viscosity ratios and/or Weber and Reynolds numbers~\cite{hu_hybrid_2019,liang_phase-field-based_2018,fakhari_weighted_2017,qin_entropic_2018}. As such we first focus on a 2-D configuration considering three sets of We and Re numbers, namely: Re=200 and We=220, Re=1000 and We=220 and Re=1000 and We=2200. In all simulations the density and viscosity ratios are set to $\rho_h/\rho_l=1000$ and $\nu_l/\nu_h=15$ emulating a water/air system. The geometrical configuration is illustrated in Figure~\ref{Fig:drop_splashing_geometry}.
The top and bottom boundary conditions are set to walls modeled with the half-way bounce-back formulation while symmetrical boundaries are applied to the left and right. The droplet diameter is resolved with 100 grid-points. The initial velocity in the droplet is set to $U_0=$0.05 and $\nu_L$ is determined via the Reynolds number:
\begin{equation}
    \hbox{Re}=\frac{\rho_h U_0 D}{\mu_h}.
\end{equation}
{Furthermore, the We number is defined as:}
\begin{equation}
{
    \hbox{We} = \frac{\rho_l D {U_0}^2}{\sigma}.
    }
\end{equation}
The evolution of the liquid surface as obtained from the simulations is shown in Figure~\ref{Fig:drop_splashing_2D_contours}. Following~\cite{josserand_droplet_2003}, breakup of the rims and splashing occurs for larger impact parameters defined as:
\begin{equation}
    K=\hbox{We}^{1/2}\hbox{Re}^{1/4}.
\end{equation}
Accordingly, the impact parameters for the studied 2-D cases are: K=55.7, 83.4 and 263.8. Looking at the evolution of the systems in Figure~\ref{Fig:drop_splashing_2D_contours} it can be clearly observed that in agreement with observations in~\cite{josserand_droplet_2003}, larger values of the impact parameter lead to droplet detachment from the rim and splashing.
Furthermore, the evolution of the spreading radii $r_K$ over time for different cases are shown in Figure~\ref{fig:drop_splashing_radius}. As shown there the radii scale with time at the initial stages of the impact, in agreement with results reported in~\cite{josserand_droplet_2003}.

\begin{figure}[!htb]
	\centering
	\hspace{-0.6\textwidth}
	\begin{subfigure}{0.3\textwidth}
		 \includegraphics{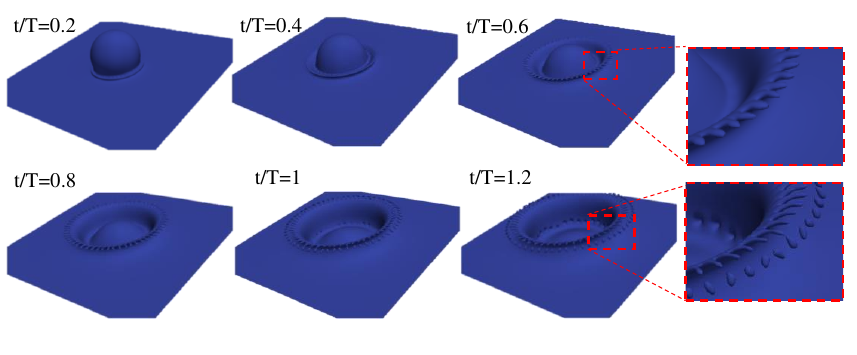}
\end{subfigure}
\caption{Impact of spherical droplet on thin liquid sheet at We=8000 and Re=1000 at different times with $\rho_h/\rho_l=1000$ and $\nu_l/\nu_h=15$.}
\label{Fig:drop_splashing_3D_contours}
\end{figure}

As a final test-case, to showcase the robustness of the proposed algorithm, a 3-D configuration with Re=1000 and We=8000 was also ran. The evolution of the liquid surface over time is shown in Figure~\ref{Fig:drop_splashing_3D_contours}.
After initial impact a thin liquid jet is formed at the contact line between the droplet and the sheet. Then, the crown evolves and spreads. At later stages the finger-like structures start to form at the tip of the crown. These liquid fingers then get detached from the crown and liquid splashing is observed. The sequence of events is in excellent agreement with those presented in~\cite{josserand_droplet_2003}. Furthermore, the spreading radius, as plotted in Figure~\ref{fig:drop_splashing_radius} is in agreement with theoretical predictions.
\section{Conclusions}
A \gls{lb}-based solver relying on the conservative \gls{ac} equation and a modified hydrodynamic pressure/velocity-based distribution and \gls{mrt} collision operator in cumulants space was presented in the this study with the aim to model multi-phase flows in the larger Weber/Reynolds regimes. While the stability at high Weber numbers --{i.e.} low surface tensions-- is achieved through the decoupled nature of the conservative \gls{ac} formulation, the added stability in terms of kinematic viscosity --{i.e.} larger Reynolds numbers-- is brought about by the collision operator and the modified pressure-based \gls{lb} formulation for the flow. {Compared to other models available in the literature based on the \gls{ac} formulation, the use of cumulants allows for stability at considerably higher Reynolds numbers --{i.e.} lower values of the relaxation factor. For instance, configurations such as the 3-D droplet splashing were not stable with the \gls{srt} formulation for the same choice of non-dimensional parameters, {i.e.} resolution and relaxation factor.} The algorithm was shown to capture the dynamics of the flow and be stable in the targeted regimes. The application of the proposed algorithm to more complex configurations such as liquid jets is currently being studied and will be reported in future publications.
\vspace{6pt} 



\authorcontributions{conceptualization, S.A.H. and H.S.; methodology, S.A.H.; software, S.A.H.; validation, S.A.H. and H.S.; formal analysis, S.A.H.; investigation, S.A.H.; data curation, S.A.H.; writing--original draft preparation, S.A.H.; writing--review and editing, S.A.H., H.S. and D.T.; visualization, S.A.H.; supervision, D.T.}

\funding{S.A.H. and H.S. would like to acknowledge the financial support of the Deutsche Forschungsgemeinschaft (DFG, German Research Foundation) in TRR 287 (Project-ID 422037413).}

\conflictsofinterest{The authors declare no conflict of interest..} 
%


\reftitle{References}


\externalbibliography{yes}
\bibliography{manuscript.bib}
\appendix
\section{Hermite polynomials and coefficients\label{App:Hermite}}
{The Hermite polynomials used in the \gls{edf}s of different solvers are defined as:}
\begin{subequations}
	\begin{align}
	\mathcal{H}_{0} &= 1, \\
	\mathcal{H}_{i} &= c_{\alpha,i}, \\
	\mathcal{H}_{ij}  &= c_{\alpha,i}c_{\alpha,j} - c_s^2\delta_{ij},
	\end{align}
\end{subequations}
{where $\delta_{ij}$ denotes the Kronecker delta function, while corresponding equilibrium coefficients are:}
\begin{subequations}
	\begin{align}
		a^{(eq)}_{0} &= \rho, \\
		a^{(eq)}_{i} &= \rho u_i, \\
		a^{(eq)}_{ij} &= \rho u_i u_j ,
	\end{align}
\end{subequations}


\end{document}